\documentclass[twocolumn,twoside]{IEEEtran}

\ifCLASSINFOpdf

\else

\fi

\ifCLASSOPTIONcompsoc
\usepackage[caption=false, font=normalsize, labelfont=sf, textfont=sf]{subfig}
\else
\usepackage[caption=false, font=normalsize]{subfig}
\fi
\usepackage{lipsum}%
\usepackage[dvipsnames]{xcolor}

\usepackage{balance}
\usepackage{multicol}   % for equalize of last page columns
\usepackage{cite}
\usepackage{gensymb}
\usepackage{multirow}
\usepackage{graphics}
\usepackage{epsfig}
\usepackage{graphicx}
\usepackage{epstopdf}
\usepackage{textcomp}
\usepackage{amsmath}
\usepackage{mathtools}
\usepackage{filecontents}
\usepackage{lipsum,color}
\usepackage{amssymb}
\usepackage{float}
\usepackage{comment}

\usepackage{amsthm}

\usepackage{braket}
\usepackage{times} % assumes new font selection scheme installed
\usepackage{amsthm}  % for proof 
\usepackage{amsfonts}

\theoremstyle{break}
\theoremstyle{break}
\begin{document}
\raggedbottom
\makeatletter
\def\fps@figure{!t} 
\def\fps@table{!t}   
\makeatother

\title{MRR-Based Line-Laser Scanning for Reliable Vehicular Positioning and Optical Communication}

\author{Mohammad~Taghi~Dabiri,~Hossein~Safi, ~{\it Member,~IEEE},~Rula~Ammuri, 
	Mazen~Hasna,~{\it Senior Member,~IEEE}, 
	~Khalid~Qaraqe,~{\it Senior Member,~IEEE},
	Harald~Haas,~{\it Fellow,~IEEE}, and Iman~Tavakkolnia,~{\it Senior Member,~IEEE}	
	
	\thanks{M.T. Dabiri and K. A. Qaraqe are with the Qatar Center for Quantum Computing, College of Science and Engineering, Hamad Bin Khalifa University, Doha, Qatar. email: (mdabiri@hbku.edu.qa; kqaraqe@hbku.edu.qa).}
	
	\thanks{M. Hasna is with the Department of Electrical Engineering, Qatar University, Doha, Qatar (e-mail: hasna@qu.edu.qa).}
		\thanks{H. safi, I. Tavakkolnia, and H. Haas are  with the LiFi R\&D centre, Department of Engineering, University of Cambridge, Cambridge, UK, e-mails: \{hs905, it360, huh21\}@cam.ac.uk.} 
}

\maketitle
%\vspace{-1cm}
%%%%%%%%%%%%%%%%%%%%%%%%%%%%%%%%%%%%%%%%%%%%%%%%%%%%%%%%%%
%%%%%%%%%%%%%%%%%%%%%%%%%%%%%%%%%%%%%%%%%%%%%%%%%%%%%%%%%%
\begin{abstract}
High-speed vehicular environments require optical systems capable of joint sensing, positioning, and communication (JSPC) without mechanical tracking. Existing optical and integrated sensing–communication approaches often rely on point-source emitters or camera-based receivers, limiting spatial coverage and update rate under highway dynamics. This work introduces a new class of tracking-free optical JSPC systems that combine structured line-laser illumination with modulating retroreflector (MRR) arrays on vehicles. Two orthogonal line lasers perform synchronized longitudinal and transverse scanning to provide continuous, wide-area coverage across the roadway. A coverage-driven analytical framework models the coupling between beam divergence, scan geometry, and dwell-time allocation, enabling joint evaluation of sensing reliability and communication quality. An optimization scheme is developed to adapt scanning and divergence parameters for uniform coverage and power efficiency. Simulation results demonstrate significant improvements in spatial coverage uniformity, link stability, and reliability within a fixed scan period. These results establish a practical pathway toward scalable, turbulence-resilient optical architectures for next-generation vehicular JSPC networks.
\end{abstract}

%
%\begin{IEEEkeywords}
%	...
%\end{IEEEkeywords}

%
\IEEEpeerreviewmaketitle

%% ---------------------------------------
%% ---------------------------------------

\section{Introduction}
Highway autonomy demands unprecedented precision, reliability, and responsiveness in vehicle perception and connectivity \cite{zhang2025vehicle,nguyen2025survey}. 
Self-driving and cooperative driving functions rely on joint positioning, sensing, and communication to achieve lane-level awareness, collision avoidance, and coordinated maneuvers at velocities often exceeding $100~\mathrm{km/h}$ \cite{decarli2023v2x, cai2023consensus}. 
Under such dynamics, centimeter-scale localization and sub-10~ms information latency are critical for maintaining safety envelopes and stable control loops. 
Conventional RF-based vehicular networks, however, face fundamental challenges in achieving such spatial and temporal precision due to bandwidth scarcity, multipath distortion, and interference congestion. 

Recent studies highlight the rapid convergence of vehicular sensing and communication technologies toward centimeter-level awareness and millisecond-scale update rates. Deep learning–based cooperative perception has shown that fusing multi-vehicle LiDAR data substantially improves localization and mapping accuracy in connected and automated vehicles (CAVs)~\cite{barbieri2024lidar}. In parallel, surveys on vehicular visible light communication (VLC) and optical wireless systems have established the physical and networking foundations enabling fine angular selectivity and high spatial resolution in vehicular environments~\cite{Memedi2021VVLC}. The emerging paradigm of integrated sensing and communication (ISAC) further demonstrates that co-designed sensing–communication frameworks can simultaneously enhance spectrum utilization and perception reliability~\cite{Liu2020ISAC,Cheng2022ISACVehicular}. Meanwhile, the recent IEEE~802.11bb standard has formalized optical PHY/MAC protocols, underscoring the technological readiness of light-based communication for vehicular-to-everything (V2X) networks~\cite{zhang2024lirf}. Despite these advances, most optical and ISAC systems rely on point-source emitters or camera-based receivers, which constrain update rates and spatial scalability under highway-speed dynamics. These limitations motivate the development of structured-beam optical architectures capable of rapid, wide-area scanning and tracking-free operation for high-speed joint sensing, positioning, and communication (JSPC).

At the link level, recent progress in retroreflective and structured optical links provides promising pathways toward realizing such architectures. Modulating retroreflectors (MRRs) have been proposed to close high-mobility free-space optical (FSO) and visible-light links by retroreflecting and modulating incident beams without vehicle-side tracking~\cite{Dabiri2022MRRFSO}. Retroreflective communication and positioning approaches exploit the geometry of the return pattern for accurate localization with minimal terminal complexity~\cite{10897286}. Complementary LiDAR-based cooperative localization studies further confirm that perception-driven methods can achieve sub-meter accuracy while reducing network overhead, making them well suited for real-time highway automation~\cite{barbieri2024lidar}.

In parallel with retroreflective techniques, line-laser architectures have recently gained attention for industrial inspection, IoT metrology, and optical ranging applications owing to their capability of generating narrow, spatially structured illumination with high scan rates and precise geometric encoding~\cite{chen2025line, ergin2024robotic}. In vehicular contexts, such line-shaped beams enable rapid surface profiling and high-speed detection of obstacles or markings without mechanical beam steering. Building on these complementary advances, this paper introduces a novel configuration that combines line-laser scanning with MRR technology to achieve joint high-precision positioning and bidirectional communication. Two orthogonal super-Gaussian line lasers, one performing longitudinal scanning and the other transverse scanning, operate synchronously to cover the full highway surface. The MRR arrays mounted on vehicle roofs retroreflect and modulate the incident optical beams, enabling simultaneous downlink sensing and uplink data transmission without the need for active tracking. This architecture supports millisecond-level position update intervals, providing continuous monitoring of high-speed vehicles with centimeter-level precision. Nevertheless, challenges such as nonlinear coverage distribution and dwell-time imbalance arise due to the geometric relationship between the azimuthal angle and longitudinal coordinate, which are analytically modeled and optimized in this work.

Building upon the above system architecture, the proposed framework establishes a new class of tracking-free optical sensing systems that integrate structured line-laser illumination with MRR-based modulation for highway-scale joint sensing, positioning, and communication. A coverage-driven analytical formulation links the received energy map to beam divergence, scan geometry, and dwell-time distribution, thereby enabling a unified assessment of sensing reliability and communication signal-to-noise ratio (SNR). Based on this model, an optimization strategy is developed to jointly adapt the azimuthal sampling parameters and divergence controls, minimizing coverage holes while preserving optical power efficiency. The resulting nonuniform scanning scheme compensates for the nonlinear mapping and yields nearly uniform longitudinal illumination across the MRR plane. Simulation results confirm that the optimized configuration increases coverage from $81.4\%$ to over $98.4\%$ (reduces the hole ratio from $18.6\%$ to $2.6\%$), without extending the $10$ ms scan period. These findings demonstrate that coverage-oriented optimization of beam geometry and scanning law can simultaneously enhance positioning accuracy and optical link stability, making the proposed design a practical enabler for next-generation high-speed vehicular JSPC systems.

%%%%%%%%%%%%%%%%%%%%%%%%%%%%%%%%%%%%%%%%%%%%%%%%%%%%%%%%%%%%%%%%
%%%%%%%%%%%%%%%%%%%%%%%%%%%%%%%%%%%%%%%%%%%%%%%%%%%%%%%%%%%%%%%% VERSUS P_T
\begin{figure}
	\begin{center}
		\includegraphics[width=3.15 in]{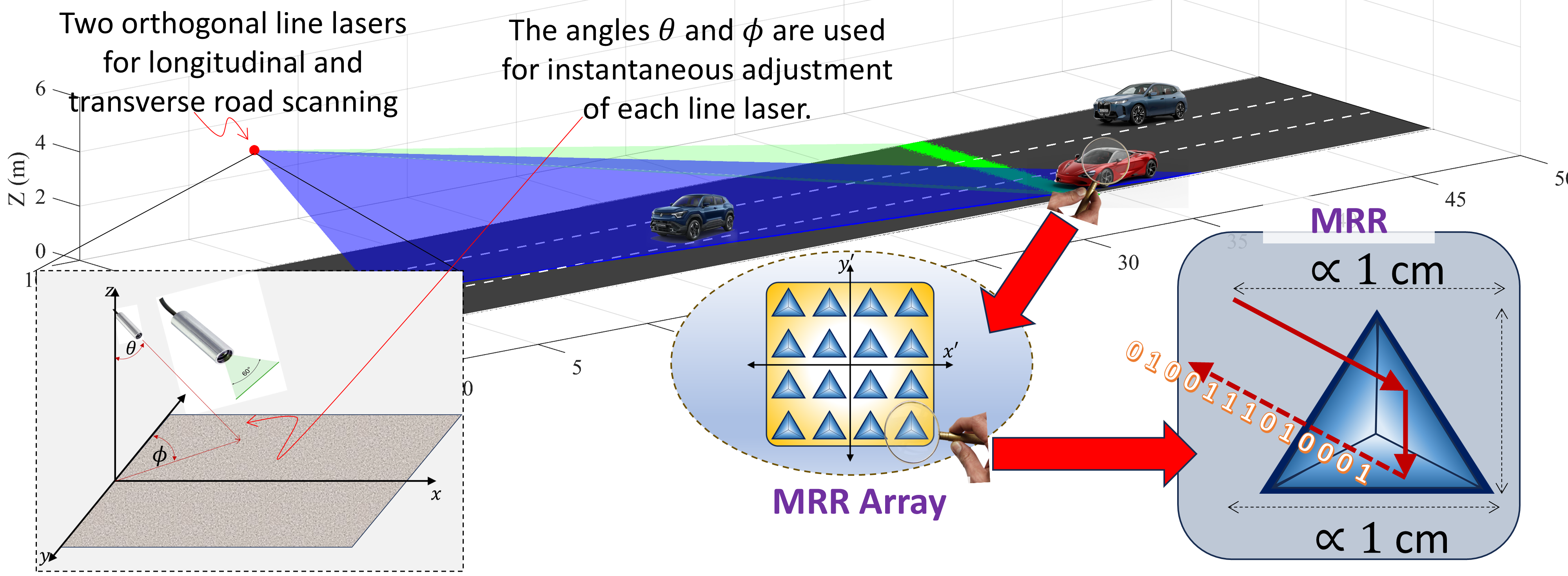}
		\caption{System model of the proposed dual-fan laser scanning configuration. 
			Two orthogonal line lasers perform longitudinal and transverse road scanning, while the instantaneous beam orientations are controlled by the elevation $\theta$ and azimuth $\phi$ angles. 
			The MRR array on vehicle roofs enables bidirectional optical links for JSPC systems.}
		
		\label{sm1}
	\end{center}
\end{figure}
%%%%%%%%%%%%%%%%%%%%%%%%%%%%%%%%%%%%%%%%%%%%%%%%%%%%%%%%%%%%%%%%
%%%%%%%%%%%%%%%%%%%%%%%%%%%%%%%%%%%%%%%%%%%%%%%%%%%%%%%%%%%%%%%%

%%-----------------------------------------------------------------
%%-----------------------------------------------------------------
\section{System Model}
\subsection{Laser Transmitter Setup}
As illustrated in Fig.~\ref{sm1}, the proposed system employs two orthogonal line lasers to perform fast longitudinal and transverse scanning over the road surface. 
The transmit unit (Tx) is mounted at height $z_{\mathrm{Tx}}$ above the road center, emitting two optical beams with independent parameters. 
Each beam, indexed by $q \in \{L, T\}$ for the longitudinal and transverse fans, respectively, is characterized by its optical power $P_{t,q}$, wavelength $\lambda_q$, and divergence parameters $(\alpha_q, \beta_q)$. 
The instantaneous orientation of each beam is controlled through the elevation and azimuth angles $(\theta_q, \phi_q)$, which are dynamically adjusted within each scanning period $T_{\mathrm{scan},q}$ to achieve full spatial coverage.

Let the transmitter position be denoted as
\begin{align}
	\mathbf{P}_0 = [x_{\mathrm{Tx}},\, y_{\mathrm{Tx}},\, z_{\mathrm{Tx}}]^{\mathrm{T}},
	\label{eq:P0}
\end{align}
where $[\cdot]^{\mathrm{T}}$ is the transpose operator. Also, let us define the unit propagation vector for beam $q$ as
\begin{align}
	\mathbf{d}_q(\theta_q,\phi_q) = 
	\begin{bmatrix}
		\sin(\theta_q)\cos(\phi_q) \\
		\sin(\theta_q)\sin(\phi_q) \\
		-\cos(\theta_q)
	\end{bmatrix}.
	\label{eq:dirq}
\end{align} 
The two orthogonal line lasers are configured such that their fan directions are mutually perpendicular on the road plane. 
For the longitudinal fan ($q=L$), the beam lies in the $x$–$z$ plane, while for the transverse fan ($q=T$) it lies in the $y$–$z$ plane. 
In the reference configuration, both beams share the same azimuth angle $\phi_q=0$, and their line-forming optics generate perpendicular scanning patterns along the $x$- and $y$-axes, respectively. 
Each beam subsequently sweeps around its central direction to achieve full road coverage.

%%%%%%%%%%%%%%%%%%%%%%%%%%%%%%%%%%%%%%%%%%%%%%%%%%%%%%%%%%%%%%%%
%%%%%%%%%%%%%%%%%%%%%%%%%%%%%%%%%%%%%%%%%%%%%%%%%%%%%%%%%%%%%%%% VERSUS P_T
\begin{figure}
	\begin{center}
		\includegraphics[width=3.15 in]{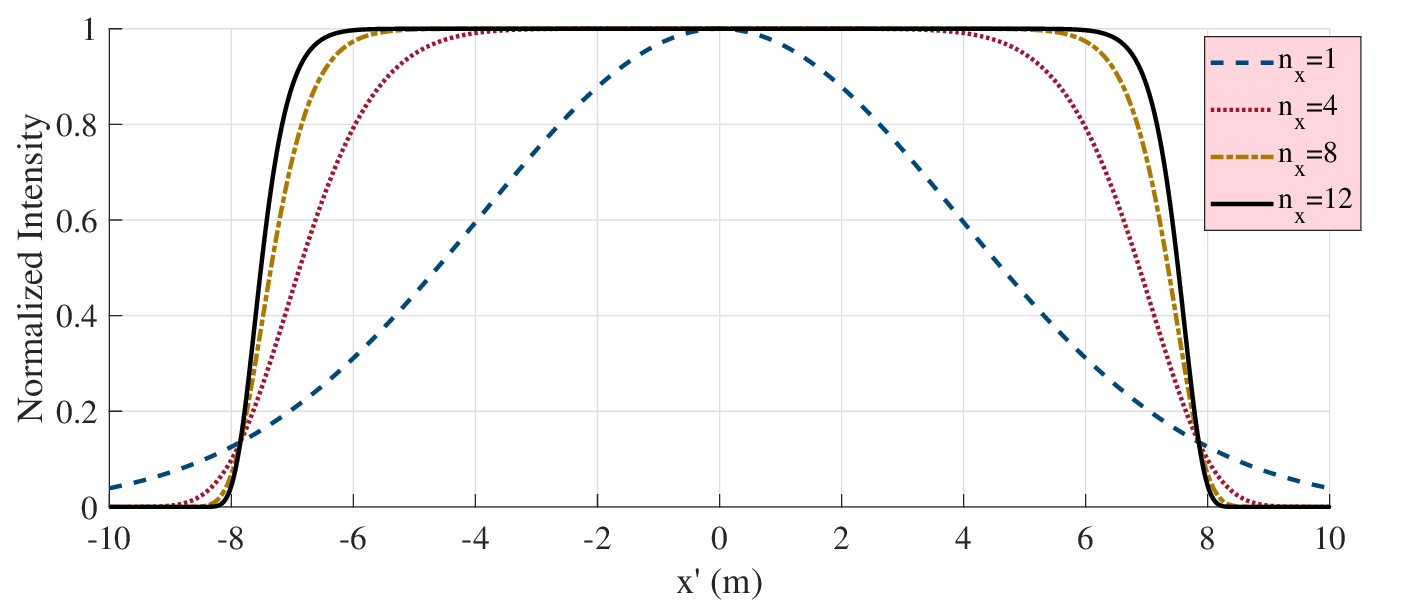}
		\caption{Comparison of super-Gaussian beam profiles along the line axis for different shape orders $n_{x,L}$. At a propagation distance of 10~m, increasing the order from $n=1$ (standard Gaussian) to $n=12$ results in a flatter and more uniform intensity distribution, approaching an ideal line-like pattern. The thickness axis remains Gaussian ($n_y=1$), while the divergence angles are set to $\theta_{\text{div},x,L}=1^\circ$ and $\theta_{\text{div},y,L}=60^\circ$.}
		\label{sm2}
	\end{center}
\end{figure}
%%%%%%%%%%%%%%%%%%%%%%%%%%%%%%%%%%%%%%%%%%%%%%%%%%%%%%%%%%%%%%%%
%%%%%%%%%%%%%%%%%%%%%%%%%%%%%%%%%%%%%%%%%%%%%%%%%%%%%%%%%%%%%%%%

The optical intensity of each beam is modeled by an anisotropic super-Gaussian distribution in the local beam coordinates $(x',y',z')$, with the propagation axis aligned to $z'$. 
The generalized intensity profile is \cite{cai2008paraxial}
\begin{align}
	&I_q(x',y',z') = I_{0,q}
	\left(\frac{w_{0x,q}}{w_{x,q}(z')}\right)
	\left(\frac{w_{0y,q}}{w_{y,q}(z')}\right) \nonumber \\[0.3em]
	&\times
	\exp\!\left[-2\!\left(
	\left|\frac{x'}{w_{x,q}(z')}\right|^{2n_{x,q}}
	+\left|\frac{y'}{w_{y,q}(z')}\right|^{2n_{y,q}}
	\right)\right],
	\label{eq:supergaussian_general}
\end{align}
where $w_{0x,q}$ and $w_{0y,q}$ denote the beam waists along the longitudinal and thickness axes, respectively.  
The exponents $n_{x,q}$ and $n_{y,q}$ define the order of the super-Gaussian distribution along each direction. Here,  
$n=1$ yields a standard Gaussian, while larger $n$ values flatten the beam profile along the corresponding axis.
The divergence angles are independent along the two axes \cite{cai2008paraxial}
\begin{align}
	\theta_{\mathrm{div},x,q} = \frac{\lambda_q}{\pi w_{0x,q}}, \qquad
	\theta_{\mathrm{div},y,q} = \frac{\lambda_q}{\pi w_{0y,q}},
	\label{eq:divergence}
\end{align}
with $\theta_{\mathrm{div},x,q}\ll\theta_{\mathrm{div},y,q}$ to achieve a narrow divergence along the line axis and a wider spread along the thickness axis. For the longitudinal fan ($q=L$), the line axis corresponds to $x'$ (\(n_{x,L}\!\gg\!1,\,n_{y,L}=1\)), 
whereas for the transverse fan ($q=T$) it corresponds to $y'$ (\(n_{x,T}=1,\,n_{y,T}\!\gg\!1\)).
Fig.~\ref{sm2} illustrates the effect of the super-Gaussian order $n_{x,L}$ on the longitudinal beam profile. 
As the order increases, the beam becomes flatter along its line axis while maintaining a narrow thickness, effectively approximating a continuous line illumination after several meters of propagation.

\subsection{MRR-based Vehicle Model}
We consider a small MRR array mounted on the vehicle roof. 
The array plane is parallel to the global $y$–$z$ plane and centered at
\begin{align}
	\mathbf{P}_{\mathrm{MRR}}=[x_v,\,y_v,\,z_v]^{\mathrm{T}}, \qquad (z_v>0),
	\label{eq:mrr_pos}
\end{align}
with physical area $A_{\mathrm{arr}}$. The MRR performs low-voltage modulation and retroreflects the incident signal back toward the Tx with overall optical efficiency $\eta_q$ (including passive reflectivity and modulator insertion efficiency) for beam.

\subsubsection{Incident Sampling on a Small MRR}
For a given orientation $(\theta_q,\phi_q)$, the beam direction is $\mathbf{d}_q$ from \eqref{eq:dirq}. 
The intersection with the MRR plane $x=x_v$ occurs at
\begin{align}
	t_q=\frac{x_v-x_{\mathrm{Tx}}}{d_{x,q}}, \qquad R_q=\|\mathbf{P}_0+t_q\mathbf{d}_q-\mathbf{P}_0\|=t_q,
	\label{eq:int_param}
\end{align}
and the incidence cosine factor (plane normal $\mathbf{n}=[1,0,0]^{\mathrm{T}}$) is
\begin{align}
	\cos\gamma_q=|\mathbf{d}_q\cdot\mathbf{n}|=|d_{x,q}|.
	\label{eq:cos_inc}
\end{align}
Let $\{\mathbf{u}_q,\mathbf{v}_q,\mathbf{d}_q\}$ be the beam-fixed orthonormal basis (fan/thickness axes and propagation). 
The beam-local coordinates of the hit point are $	z'_q=t_q$ and
\begin{align} 
	x'_q=(\mathbf{P}_0+t_q\mathbf{d}_q-\mathbf{P}_0)\!\cdot\!\mathbf{u}_q,\quad
	y'_q=(\mathbf{P}_0+t_q\mathbf{d}_q-\mathbf{P}_0)\!\cdot\!\mathbf{v}_q.
	\label{eq:local_xy}
\end{align}
Under the {small-aperture assumption} (i.e., when intensity is nearly constant over the array), the incident power on the MRR is
\begin{align}
	P^{(\mathrm{inc})}_{q} \;=\; I_q\!\big(x'_q,y'_q,z'_q\big)\; A_{\mathrm{arr}}\; \cos\gamma_q,
	\label{eq:Pinc_simple}
\end{align}
where $I_q(\cdot)$ is the beam profile in \eqref{eq:supergaussian_general}.

\subsubsection{Received Power at the Co-Located Receiver}
The retroreflected signal is directed back toward Tx. With overall MRR efficiency $\eta_q$, the received power collected at the co-located aperture is modeled as
\begin{align}
	P^{(\mathrm{rx})}_{q} \;=\; \eta_q\,\kappa_q(R_q)\, P^{(\mathrm{inc})}_{q},
	\label{eq:Prx_simple}
\end{align}
where $\kappa_q(R_q)\!\in\!(0,1]$ is a lumped capture factor accounting for collection by the receive optics (e.g., limited aperture and retro-lobe width). 
In the simplest aperture-limited model with receiver area $A_{\mathrm{rx}}$ and retro-lobe half-angle $\delta_q$, we have 
\begin{align}
	\kappa_q(R_q)\;\approx\;\min\!\left\{1,\; \frac{A_{\mathrm{rx}}}{\pi\big(R_q\,\delta_q\big)^2}\right\}.
	\label{eq:kappa_optional}
\end{align}
Consequently, \eqref{eq:Pinc_simple} and \eqref{eq:Prx_simple} provide a closed-form, traceable link budget from the beam parameters $(\theta_q,\phi_q)$ and vehicle position $(x_v,y_v,z_v)$ to the received power, enabling direct evaluation of $P^{(\mathrm{rx})}_q$ for any instantaneous vehicle location and beam orientation. This received-power model serves as the foundation for the coverage and SNR analyses presented in the next section.

%%%%%%%%%%%%%%%%%%%%%%%%%%%%%%%%%%%%%%%%%%%%%%%%%%%%%%%%%%%%%%%%
%%%%%%%%%%%%%%%%%%%%%%%%%%%%%%%%%%%%%%%%%%%%%%%%%%%%%%%%%%%%%%%% VERSUS W_Z
\begin{figure*}
	\centering
	\subfloat[] {\includegraphics[width=6.2 in]{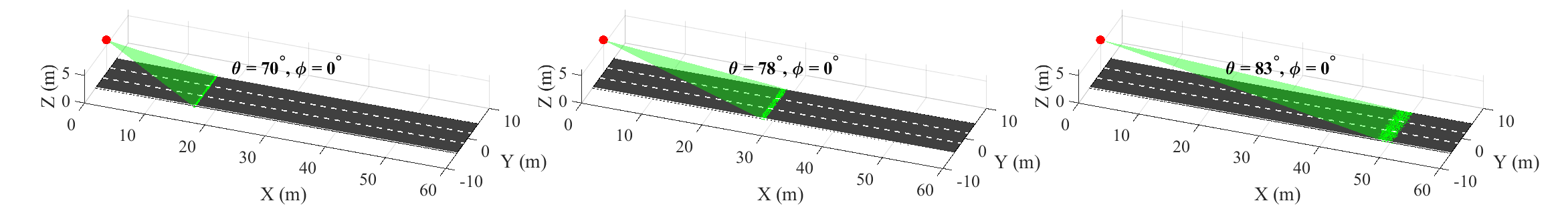}
		\label{cz1}
	}
	\hfill
	\subfloat[] {\includegraphics[width=6.2 in]{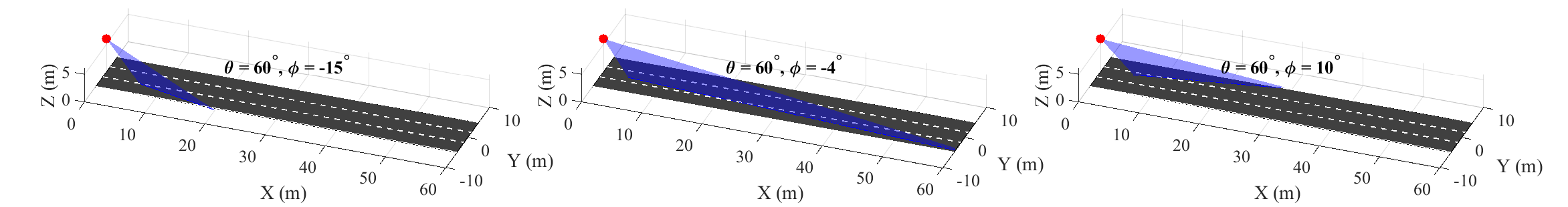}
		\label{cz2}
	}	
	\caption{Three-dimensional visualization of the beam footprints for the two orthogonal scanning fans: (a) illustrates the {transverse fan}, where variations in the elevation angle $\theta_T$ steer the beam across the road width, 
		while (b) shows the {longitudinal fan}, where changes in the azimuth angle $\phi_L$ sweep along the road length. 
		Both fans are generated by line-shaped beams with anisotropic super-Gaussian profiles defined by $(n_{x,q},n_{y,q})=(1,8)$ or $(8,1)$, and divergence angles $(\theta_{x,q},\theta_{y,q})=(1^{\circ},60^{\circ})$, illustrating the orthogonal scanning coverage geometry of the proposed system.}
	
	\label{cz}
\end{figure*}
%%%%%%%%%%%%%%%%%%%%%%%%%%%%%%%%%%%%%%%%%%%%%%%%%%%%%%%%%%%%%%%%
%%%%%%%%%%%%%%%%%%%%%%%%%%%%%%%%%%%%%%%%%%%%%%%%%%%%%%%%%%%%%%%% 
%% ------------------------------------------------------------
%% ------------------------------------------------------------
\section{Scanning Mechanism}
Each of the two line lasers performs a periodic angular sweep to ensure full coverage of the road surface. 
The scanning pattern for the longitudinal fan and the transverse fan follows distinct angular control laws corresponding to their geometrical orientation.

%-------------------------------------
\subsection{Transverse Fan (Y-direction Sweep)}
The transverse fan keeps a fixed azimuth angle $\phi_T = 0$ and performs a sweep over the elevation angle $\theta_T$, covering the road width. 
The discrete elevation angles are defined as
\begin{align}
	\theta_{T,k} \in [\theta_{\min,T},\,\theta_{\max,T}], 
	\label{eq:theta_sweep}
\end{align}
which produces a narrow illuminated line across the road in the $y$-direction.

We set the coverage at the MRR height, not on the road plane. 
Let $z_{\mathrm{MRR}}\!\in(0,z_{\mathrm{Tx}})$ denote the design elevation at which the transverse fan must span the full road width.\footnote{Typical passenger-vehicle roof heights fall well below $z_{\mathrm{Tx}}$; the design equations below remain valid for any $z_{\mathrm{MRR}}$.}
With fixed $\phi_T\!=\!0$, the central ray direction is $\mathbf{d}_T=[\sin\theta_T,\,0,\,-\cos\theta_T]^{\mathrm{T}}$ as in \eqref{eq:dirq}. 
The intersection with the plane $z=z_{\mathrm{MRR}}$ occurs at the (positive) range parameter
\begin{align}
	s_T(\theta_T)=\frac{z_{\mathrm{Tx}}-z_{\mathrm{MRR}}}{\cos\theta_T}.
	\label{eq:sT}
\end{align}
Also, let $\mathbf{u}_T$ be the lateral unit vector of the transverse beam (fan axis), thus, for $\phi_T=0$ we have $\mathbf{u}_T=[0,1,0]^{\mathrm{T}}$.
The boundary rays are formed by adding a symmetric half-angle $\tfrac{\theta_{\mathrm{div},y,T}}{2}$ in the $\pm\mathbf{u}_T$ direction, i.e., we have
\begin{equation}
\mathbf{d}^{(\pm)}=\frac{\mathbf{d}_T+\tan(\tfrac{\theta_{\mathrm{div},y,T}}{2})\,(\pm\mathbf{u}_T)}
{\big\|\mathbf{d}_T+\tan(\tfrac{\theta_{\mathrm{div},y,T}}{2})\,(\pm\mathbf{u}_T)\big\|}.
\end{equation}
At $z=z_{\mathrm{MRR}}$, their lateral coordinates are
\begin{align}
	y_T^{(\pm)}(\theta_T)
	= \pm\frac{z_{\mathrm{Tx}}-z_{\mathrm{MRR}}}{\cos\theta_T}\,
	\tan\!\Big(\tfrac{\theta_{\mathrm{div},y,T}}{2}\Big).
	\label{eq:yTpm}
\end{align}
Therefore, the illuminated span on the MRR plane is
\begin{align}
	W_T^{(\mathrm{MRR})}(\theta_T)
	\;=\; \frac{2\,(z_{\mathrm{Tx}}-z_{\mathrm{MRR}})}{\cos\theta_T}\,
	\tan\!\Big(\tfrac{\theta_{\mathrm{div},y,T}}{2}\Big).
	\label{eq:W_T_MRR}
\end{align}
Imposing $W_T^{(\mathrm{MRR})}(\theta_T)\!\geq\! W_{\mathrm{road}}$ yields the {required} fan divergence to be as
\begin{align}
	\theta_{\mathrm{div},y,T}^{\star}(\theta_T)
	\;=\; 2\,\tan^{-1}\!\!\left(
	\frac{W_{\mathrm{road}}\cos\theta_T}{2\,(z_{\mathrm{Tx}}-z_{\mathrm{MRR}})}
	\right),
	\label{eq:Theta_y_req}
\end{align}
which is monotonically decreasing with $\theta_T$.
In practice we set $\theta_{\mathrm{div},y,T}=\min\{\theta_{\mathrm{div},y,\max},\,\theta_{\mathrm{div},y,T}^{\star}(\theta_T)\}$ to respect the optics/mechanics limits.

The thickness half-angle $\tfrac{\theta_{\mathrm{div},x,T}}{2}$ tilts the beam within the $(x,z)$ plane (basis vector $\mathbf{v}_T$), producing a finite vertical spread at $z=z_{\mathrm{MRR}}$. 
Following the same geometric formulation, the footprint width along the $x$-axis is obtained as
\begin{align}
	W_{x,T}^{(\mathrm{MRR})}(\theta_T)
	\;=\; \frac{2\,(z_{\mathrm{Tx}}-z_{\mathrm{MRR}})\,
		\tan\!\big(\tfrac{\theta_{\mathrm{div},x,T}}{2}\big)\,\sec^{2}\!\theta_T}
	{1-\tan^{2}\!\big(\tfrac{\theta_{\mathrm{div},x,T}}{2}\big)\tan^{2}\!\theta_T},
	\label{eq:Wx_MRR_exact}
\end{align}
which for small divergences simplifies to
\begin{align}
	W_{x,T}^{(\mathrm{MRR})}(\theta_T)
	\approx 2\,(z_{\mathrm{Tx}}-z_{\mathrm{MRR}})\,
	\tan\!\big(\tfrac{\theta_{\mathrm{div},x,T}}{2}\big)\sec^{2}\!\theta_T.
	\label{eq:Wx_MRR_simplified}
\end{align}
Figure~\ref{cz1} illustrates the angular scanning of the transverse fan along the $y$-direction, where the beam orientation $\theta_T$ is varied to sweep the coverage across the road width at the MRR height. 
As $\theta_T$ increases, the beam footprint expands in both the longitudinal and vertical directions where (i) the effective path length $s_T(\theta_T)$ in \eqref{eq:sT} increases, and 
(ii) the projection factor $\sec^{2}\!\theta_T$ amplifies the apparent divergence.
Therefore, the illuminated region becomes wider and slightly thicker at higher elevation angles, as visualized in Fig.~\ref{cz1} for three representative values of~$\theta_T$.

%%------------------------------------------------------
%-------------------------------------
\subsection{Longitudinal Fan (X-direction Sweep)}
The longitudinal fan keeps a fixed elevation $\theta_L$ and performs an azimuthal sweep over $\phi_L$ to cover the road length. 
The discrete azimuth angles are
\begin{align}
	\phi_{L,k} \in [\phi_{\min,L},\,\phi_{\max,L}], 
	\label{eq:phi_sweep}
\end{align}
which produces a narrow illuminated line along the $x$-direction.
We again set the coverage at the MRR height $z=z_{\mathrm{MRR}}$. 
With central direction $\mathbf{d}_L=[\sin\theta_L\cos\phi_L,\,\sin\theta_L\sin\phi_L,\,-\cos\theta_L]^{\mathrm{T}}$ \eqref{eq:dirq}, 
the intersection range is independent of $\phi_L$. Thus, we have
\begin{align}
	r_L \triangleq s_L\sin\theta_L=(z_{\mathrm{Tx}}-z_{\mathrm{MRR}})\tan\theta_L ,
	\label{eq:sL_rL}
\end{align}
where $s_L=\frac{z_{\mathrm{Tx}}-z_{\mathrm{MRR}}}{\cos\theta_L}$. Therefore, the footprint center at $z_{\mathrm{MRR}}$ traces the locus 
$x=r_L\cos\phi_L,\ y=r_L\sin\phi_L$ on the $(x,y)$-plane.

Let the symmetric azimuthal spread around $\phi_L$ be $\Delta\phi$. 
Feasibility at $z=z_{\mathrm{MRR}}$ requires the two edge rays $\phi_L\pm\Delta\phi$ to satisfy the road constraints 
$|y|\leq W_{\mathrm{road}}/2$ and $0\leq x\leq L_{\mathrm{road}}$. 
The lateral constraint gives the dominant bound $	|\sin\phi|\leq \frac{W_{\mathrm{road}}}{2r_L}$ for $\phi\in[-\phi_{\lim},\,\phi_{\lim}]$, then
\begin{align}
	\phi_{\lim}=\sin^{-1}\!\!\left(\min\!\left\{1,\frac{W_{\mathrm{road}}}{2r_L}\right\}\right).
	\label{eq:phi_limit}
\end{align}
Hence, the admissible half–spread is $\Delta\phi_{\max}(\phi_L)=\max\{0,\,\phi_{\lim}-|\phi_L|\}$, and the {required} line–axis divergence is obtained as
\begin{align}
	&\theta_{\mathrm{div},x,L}^{\star}(\phi_L)
	=\; 2\,\Delta\phi_{\max}(\phi_L) \nonumber \\
	%------------
	&~~~=\; 2\max\!\left\{0,\,
	\sin^{-1}\!\!\left(\min\!\left\{1,\frac{W_{\mathrm{road}}}{2r_L}\right\}\right)-|\phi_L|
	\right\}.
	\label{eq:theta_div_x_req}
\end{align}
In practice, we set 
$\theta_{\mathrm{div},x,L}=\min\{\theta_{\mathrm{div},x,\max},\,\theta_{\mathrm{div},x,L}^{\star}(\phi_L)\}$, 
which mirrors the simulation policy where the fan spread is twice the largest feasible $|\Delta\phi|$ that keeps both edges within the road bounds at $z=z_{\mathrm{MRR}}$.
Figure~\ref{cz2} illustrates the azimuthal scanning of the longitudinal fan along the $x$-direction at three representative values of $\phi_L$. 
As $|\phi_L|$ increases, the admissible spread in \eqref{eq:theta_div_x_req} shrinks due to the lateral bound in \eqref{eq:phi_limit}.

%% ----------------------------------------------------------
%%------------------------------------------------------------
%% ----------------------------------------------------------
%%------------------------------------------------------------
\section{Coverage Analysis}
The reliability of positioning, sensing, and communication functionalities in the proposed system is fundamentally limited by the received optical signal power at the MRR array.
Each application operates with a specific sensitivity threshold, denoted by $\Gamma_{\mathrm{pos}}$, $\Gamma_{\mathrm{sen}}$, and $\Gamma_{\mathrm{com}}$, respectively.
For reliable operation, the instantaneous received power at the MRR, i.e., $P_q^{(\mathrm{rx})}(x,y)$ in \eqref{eq:Prx_simple}, must exceed the relevant threshold at least once during the full scanning cycle.
Hence, a {coverage metric} is defined as the spatial probability or area fraction over the road where this condition is met.

At any point $(x,y)$ on the MRR plane $z=z_{\mathrm{MRR}}$, the received power from beam $q\!\in\!\{L,T\}$ is
\begin{align}
	&P_q^{(\mathrm{rx})}(x,y)
	=\eta_q\,\kappa_q(R_q)\nonumber \\
	&~~~\times I_q\!\big(x'_q(x,y),y'_q(x,y),z'_q(x,y)\big)\,A_{\mathrm{arr}}\cos\gamma_q,
	\label{eq:Prx_cov}
\end{align}
where $I_q(\cdot)$ follows the super-Gaussian intensity in \eqref{eq:supergaussian_general}.
Furthermore, during each scanning operation, the beam orientation is held quasi-static within small dwell intervals indexed by $k$.
For each angular state $(\theta_{q,k},\phi_{q,k})$, the received signal at the MRR corresponds to a single spatial illumination pattern. 
For each dwell interval associated with angular state $(\theta_{q,k},\phi_{q,k})$, 
the instantaneous received energy at $(x,y)$ can be integrated over its local time span $\tau_{q,k}$ as
\begin{align}
	E_q(x,y\!\mid\!\theta_{q,k},\phi_{q,k})
	&=\!\!\int_{t_k}^{t_k+\tau_{q,k}}\!\!\!
	P_q^{(\mathrm{rx})}\!\big(x,y,t\mid\theta_{q,k},\phi_{q,k}\big)\,dt \nonumber \\
	%---------------------------------
	&\;\approx\;
	P_q^{(\mathrm{rx})}(x,y\!\mid\!\theta_{q,k},\phi_{q,k})\,\tau_{q,k},
	\label{eq:E_subinterval}
\end{align}
where the approximation holds since $(\theta_q,\phi_q)$ vary negligibly within $\tau_{q,k}<<1$.
Accordingly, the localized energy map $E_q(x,y\!\mid\!\theta_{q,k},\phi_{q,k})$ in \eqref{eq:E_subinterval} defines the instantaneous received field distribution for each scan state. 
For a given threshold $\Gamma_{\ell}$, associated with an application layer $\ell\!\in\!\{\mathrm{pos},\mathrm{sen},\mathrm{com}\}$, 
the spatial coverage region at that orientation is defined as
\begin{align}
	\mathcal{C}_{q,\ell}(\theta_{q,k},\phi_{q,k})
	=\{(x,y):E_q(x,y\!\mid\!\theta_{q,k},\phi_{q,k})\geq\Gamma_{\ell}\}.
	\label{eq:cov_region}
\end{align}
The instantaneous coverage ratio for that state is defined as the normalized area of $\mathcal{C}_{q,\ell}$ over the full scanning domain $\mathcal{A}$
\begin{align}
	\rho_{q,\ell}(\theta_{q,k},\phi_{q,k})
	=\frac{|\mathcal{C}_{q,\ell}(\theta_{q,k},\phi_{q,k})|}{|\mathcal{A}|},
	\label{eq:cov_ratio}
\end{align}
where $|\cdot|$ denotes the spatial area measure on the MRR plane.

For an entire scanning sequence with $K_q$ angular samples, 
the effective coverage metric of beam $q$ at level $\Gamma_{\ell}$ is represented as the union of all illuminated subsets
\begin{align}
	\mathcal{U}_{q,\ell}
	=\bigcup_{k=1}^{K_q}\mathcal{C}_{q,\ell}(\theta_{q,k},\phi_{q,k}),\qquad
	\rho_{q,\ell}^{(\mathrm{eff})}
	=\frac{|\mathcal{U}_{q,\ell}|}{|\mathcal{A}|}.
	\label{eq:cov_union}
\end{align}
The value $\rho_{q,\ell}^{(\mathrm{eff})}$ therefore quantifies the spatial reliability of the scanning process for the specified threshold and beam type.

Also, to incorporate both scanning fans, two composite coverage metrics are defined as
	\begin{align}
		\mathcal{U}_{\ell}^{(\wedge)}
		&=\mathcal{U}_{L,\ell}\cap\mathcal{U}_{T,\ell},
		&\rho_{\ell}^{(\wedge)}
		&=\frac{|\mathcal{U}_{L,\ell}\cap\mathcal{U}_{T,\ell}|}{|\mathcal{A}|},\label{eq:cov_joint_and}\\
		\mathcal{U}_{\ell}^{(\vee)}
		&=\mathcal{U}_{L,\ell}\cup\mathcal{U}_{T,\ell},
		&\rho_{\ell}^{(\vee)}
		&=\frac{|\mathcal{U}_{L,\ell}\cup\mathcal{U}_{T,\ell}|}{|\mathcal{A}|},\label{eq:cov_joint_or}
	\end{align}
	where $\wedge$ and $\vee$ denote the logical intersection (joint coverage required, e.g., for positioning) and union (either-beam coverage sufficient, e.g., for communication) operations, respectively.
For analytical or simulation-based evaluation, 
	the continuous spatial domain $\mathcal{A}$ is discretized into finite cells of dimensions $\Delta x\times\Delta y$, 
	and the coverage fraction is approximated as
	\begin{align}
		&\rho_{q,\ell}(\theta_{q,k},\phi_{q,k})
		\approx
		\frac{1}{N_x N_y} \nonumber \\
		%-------------------------
		&~~~\times \sum_{i=1}^{N_x}\sum_{j=1}^{N_y}
		\mathbf{1}\!\left\{E_q(x_i,y_j\!\mid\!\theta_{q,k},\phi_{q,k})\geq\Gamma_{\ell}\right\},
		\label{eq:cov_discrete}
	\end{align}
	where $\mathbf{1}\{\cdot\}$ is the indicator function and $N_x$, $N_y$ denote the number of spatial samples along $x$ and $y$.

\subsection{Nonuniform Azimuthal Scanning Model for Longitudinal Coverage}
In the longitudinal fan, the beam azimuth $\phi_L$ determines the spatial projection of the illumination along the road axis.
Because the mapping $x=r_L\cos\phi_L$ with $r_L=(z_{\mathrm{Tx}}-z_{\mathrm{MRR}})\tan\theta_L$ is highly nonlinear, 
a uniform azimuthal progression produces a nonuniform sweep velocity across $x$, 
resulting in uneven optical dwell and inconsistent coverage density. 
The effective spatial rate of the scanning footprint satisfies
\begin{align}
	\frac{dx}{d\phi_L}=-r_L\sin\phi_L,
	\label{eq:dx_dphi_nonuniform}
\end{align}
indicating that near $\phi_L=0$ the beam moves slowly, 
while near the angular edges ($|\phi_L|\!\to\!\phi_{\max}$) it travels rapidly, 
reducing the local exposure time. 
To mitigate this imbalance, the azimuthal sequence is defined using a controlled, nonuniform angular spacing.

Accordingly, let $\Delta\phi_0>0$ denote the initial angular step around the center, 
and $\alpha>1$ the geometric expansion ratio governing the step growth away from $\phi_L=0$.
The sequence of positive azimuthal nodes is recursively defined as
\begin{align}
	\phi^{(+)}_0&=0,\quad
	\phi^{(+)}_m=\phi^{(+)}_{m-1}+\Delta\phi_{m-1},\quad
	\Delta\phi_m=\alpha\,\Delta\phi_{m-1},
	\label{eq:phi_recurrence}
\end{align}
which yields the closed-form expression
\begin{align}
	\phi^{(+)}_m=\Delta\phi_0\,\frac{\alpha^{m}-1}{\alpha-1},\qquad
	m=1,2,\ldots,M,
	\label{eq:phi_closed_form}
\end{align}
where $M$ is the largest index satisfying $\phi^{(+)}_M\le\phi_{\max}$.
The complete scanning grid is then obtained by symmetry as
\begin{align}
	\Phi_{L}
	=\{-\phi^{(+)}_M,\ldots,-\phi^{(+)}_1,0,\phi^{(+)}_1,\ldots,\phi^{(+)}_M\},
	\label{eq:phi_sweep_full_model}
\end{align}
where $K_L=2M+1$, and $K_L$ is the total number of azimuthal states during one scanning cycle.
Each azimuthal state $(\theta_L,\phi_{L,k})$ is allocated a dwell duration $\tau_{L,k}$ proportional to its angular span as
\begin{align}
	\tau_{L,k}
	=\frac{\Delta\phi_{L,k}}{\sum_{i=1}^{K_L}\Delta\phi_{L,i}}\,T_{\mathrm{scan},L},
	\label{eq:dwell_distribution}
\end{align}
ensuring a constant angular rate $\dot{\phi}_L$ but a variable effective dwell per spatial location due to the nonlinear mapping \eqref{eq:dx_dphi_nonuniform}.

\paragraph*{Parameter dependency and optimality}
The pair $(\Delta\phi_0,\alpha)$ determines both the angular sampling density and the spatial dwell distribution. 
Given $\phi_{\max}$, the number of azimuthal states is
\begin{align}
	M=\Bigg\lfloor\frac{\ln\!\Big(1+\tfrac{(\alpha-1)\phi_{\max}}{\Delta\phi_0}\Big)}{\ln\alpha}\Bigg\rfloor,\qquad K_L=2M+1,
	\label{eq:M_derived}
\end{align}
which directly affects $\tau_{L,k}$ through \eqref{eq:dwell_distribution}.
Optimal parameter selection is {topology-dependent}, governed by road geometry ($W_{\mathrm{road}},L_{\mathrm{road}}$) and transmitter height $(z_{\mathrm{Tx}},z_{\mathrm{MRR}})$ as 
\begin{align}
	&\max_{\Delta\phi_0,\alpha}\ 
	\rho_{L,\ell}^{(\mathrm{eff})} \nonumber \\
	%--------------------
	&\text{s.t.}\quad
	\phi^{(+)}_M\le\phi_{\max},\ \
	\theta_{\mathrm{div},x,L}(\phi_{L,k})\le\theta_{\mathrm{div},x,\max}.
	\label{eq:opt_phiL}
\end{align}
Small $\Delta\phi_0$ and moderate $\alpha>1$ yield a dense sampling near the center and gradually wider spacing toward the edges, 
compensating for the nonlinear Jacobian $|dx/d\phi_L|$ and leading to uniform longitudinal coverage across the MRR plane.

%%%%%%%%%%%%%%%%%%%%%%%%%%%%%%%%%%%%%%%%%%%%%%%%%%%%%%%%%%%%%%%%
%%%%%%%%%%%%%%%%%%%%%%%%%%%%%%%%%%%%%%%%%%%%%%%%%%%%%%%%%%%%%%%% VERSUS W_Z
\begin{figure}
	\centering
	\subfloat[] {\includegraphics[width=3.3 in]{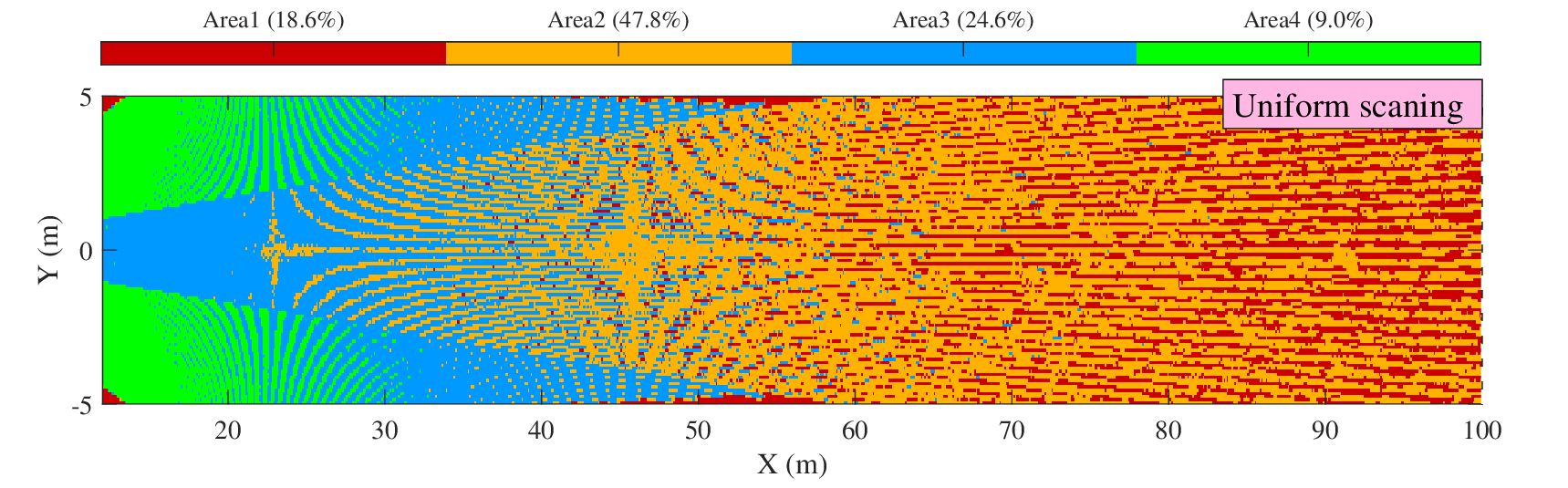}
		\label{cb1}
	}
\vspace{-.4cm}
	\hfill
	\subfloat[] {\includegraphics[width=3.3 in]{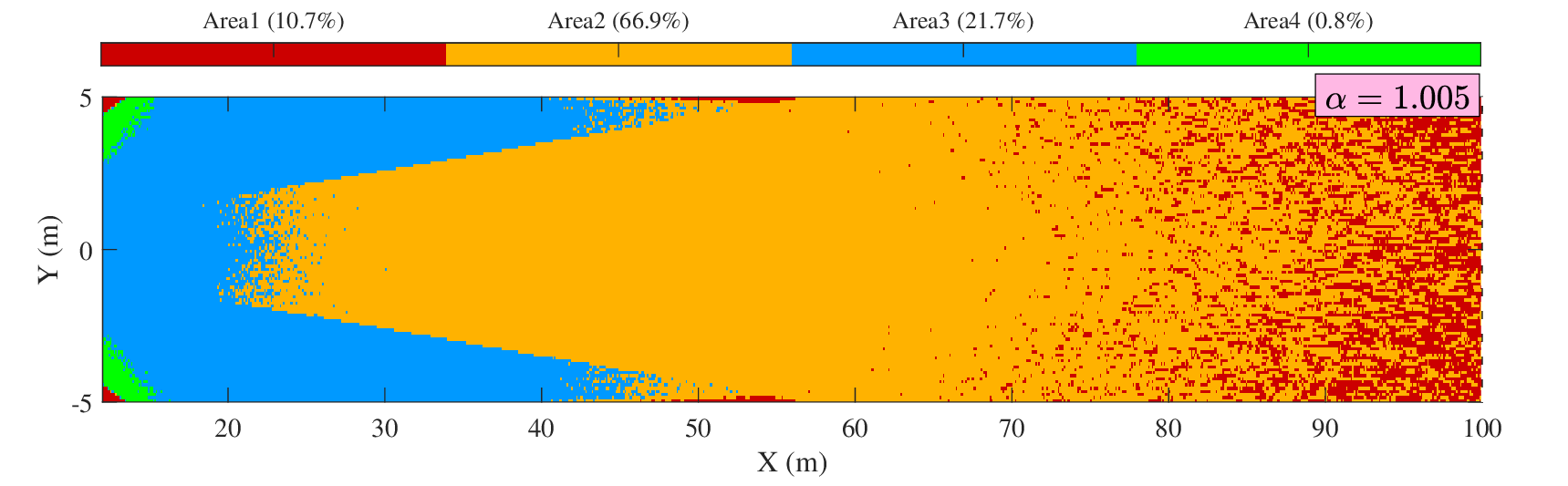}
		\label{cb2}
	}
\vspace{-.4cm}	
	\hfill
	\subfloat[] {\includegraphics[width=3.3 in]{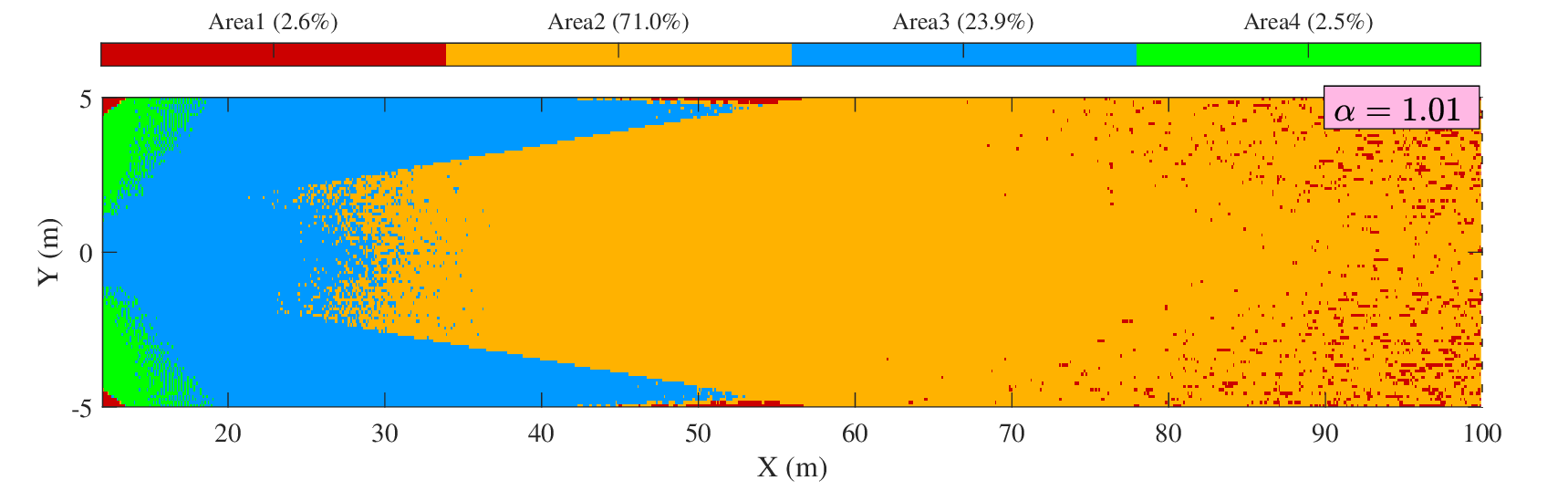}
		\label{cb3}
	}
\vspace{-.4cm}
	\hfill
	\subfloat[] {\includegraphics[width=3.3 in]{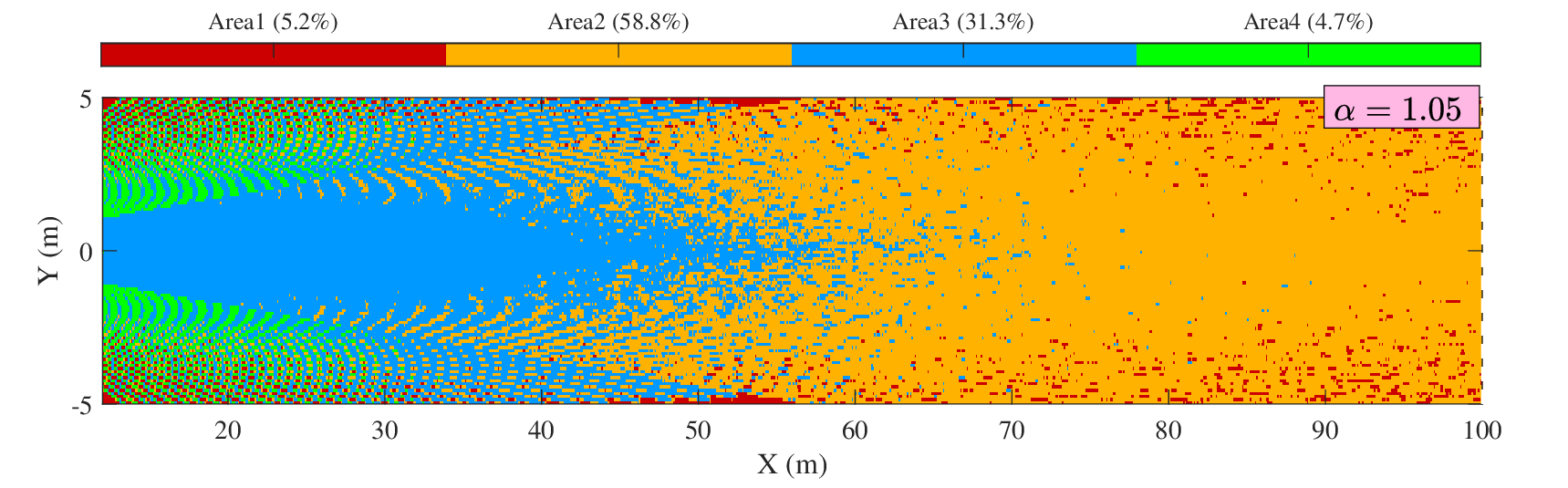}
		\label{cb4}
	}
\vspace{-.2cm}
	\caption{Longitudinal coverage maps for different azimuthal expansion factors(a) uniform scanning; (b) $\alpha=1.005$, (c) $\alpha=1.01$; and (d) $\alpha=1.05$.
		\emph{Area~1} indicates coverage holes (no precise positioning), 
		\emph{Area~2} enables centimetre-level positioning, 
		and \emph{Areas~3–4} support joint positioning and communication up to $5$ Mb/s and $50$ Mb/s, respectively. }
		%Uniform scanning ($\alpha\!=\!1$) yields strong nonuniformity, while moderate nonuniformity ($\alpha\!\approx\!1.01$) provides the most balanced and homogeneous coverage distribution.}
	\label{cb}
\end{figure}
%%%%%%%%%%%%%%%%%%%%%%%%%%%%%%%%%%%%%%%%%%%%%%%%%%%%%%%%%%%%%%%%
%%%%%%%%%%%%%%%%%%%%%%%%%%%%%%%%%%%%%%%%%%%%%%%%%%%%%%%%%%%%%%%% 

The coverage performance of the proposed nonuniform azimuthal scanning model was evaluated for a $100~\mathrm{m}$ highway segment of width $10~\mathrm{m}$, with transmitter height $z_{\mathrm{Tx}}\!=\!6.5~\mathrm{m}$ and MRR plane height $z_{\mathrm{MRR}}\!=\!1.5~\mathrm{m}$. 
The laser parameters were $P_t\!=\!0.5~\mathrm{W}$, $\lambda\!=\!1550~\mathrm{nm}$, and $T_{\mathrm{scan},L}\!=\!10~\mathrm{ms}$.
The azimuth range was $\phi_{\max}\!=\!20^{\circ}$ with $\theta_L\!=\!60^{\circ}$, and the initial step and expansion factor were varied as $(\Delta\phi_0,\alpha)\!\in\!\{0.02^{\circ},[1,1.005,1.01,1.05]\}$.

Fig.~\ref{cb} compares longitudinal coverage under a uniform sweep and three nonuniform azimuthal samplings. 
Four annotated regions are shown: {Area~1} (red) denotes coverage holes where the received energy is insufficient for precise positioning; {Area~2} marks zones enabling centimetre-level positioning within the scan window; {Areas~3–4} indicate regimes where the reflected signal supports joint communication (with {Area~4} sustaining up to $50$\,Mb/s, while {Area~3} corresponds to a lower data-rate regime, e.g., $5$\,Mb/s). 
The hole fraction ({Area~1}) decreases markedly as the azimuthal expansion factor increases from uniform to moderately nonuniform sampling, i.e.,
$18.6\%$ at $\alpha\!=\!1$ (worst case), 
$10.7\%$ at $\alpha\!=\!1.005$, 
and a minimum of $2.6\%$ at $\alpha\!=\!1.01$. For stronger stretching $\alpha\!=\!1.05$, holes slightly rise to $5.2\%$, reflecting over-compensation at the angular edges. 
Concurrently, the distributions of {Area~2} (high-precision positioning) and {Areas~3–4} (communication-capable zones) vary with $\alpha$, where moderate nonuniformity ($\alpha\approx1.01$) provides the best trade-off between hole suppression and the expansion of positioning/communication coverage. In contrast, larger $\alpha$ values further enhance baseline coverage but reduce the extent of the highest-intensity {Area~4}. Therefore, $\alpha$ should be jointly optimized with $(\Delta\phi_0,\phi_{\max})$ to maximize application-specific objectives, whether emphasizing pure positioning or joint positioning and communication performance at $5$ or $50$ Mb/s.

All numerical results presented in Fig.~\ref{cb} are obtained based on the analytical framework 
developed in \eqref{eq:Prx_cov}--\eqref{eq:opt_phiL}, which is consistent with experimentally validated 
models reported in prior optical and MRR-based studies. This ensures a direct linkage between the 
theoretical formulation and the simulated coverage performance.

\vspace{-3mm}
\section*{Acknowledgment}
\small{This publication was made possible by NPRP14C-0909-210008 from the Qatar Research, Development and Innovation (QRDI) Fund (a member of Qatar Foundation). The statements made herein are solely the responsibility of the authors.}

\small{The contributions of Dr. Hossein Safi, Dr. Iman Tavakkolnia, and Prof. Harald Haas were supported by  the Engineering and Physical Sciences Research Council (EPSRC) under grants EP/X04047X/1 and EP/Y037243/1 (``TITAN'').}

%\section{Conclusion and Future Work}

\vspace{-2mm}
%%%%%%%%%%%%%%%%%%%%%%%%%%%%%%%%%%%%%%%%%%%%%%%%%%%%%%%%%%%%%%
%%%%%%%%%%%%%%%%%%%%%%%%%%%%%%%%%%%%%%%%%%%%%%%%%%%%%%%%%%%%%%
\bibliographystyle{IEEEtran}
%\balance
\bibliography{IEEEabrv,myref}

\begin{thebibliography}{10}
\providecommand{\url}[1]{#1}
\csname url@samestyle\endcsname
\providecommand{\newblock}{\relax}
\providecommand{\bibinfo}[2]{#2}
\providecommand{\BIBentrySTDinterwordspacing}{\spaceskip=0pt\relax}
\providecommand{\BIBentryALTinterwordstretchfactor}{4}
\providecommand{\BIBentryALTinterwordspacing}{\spaceskip=\fontdimen2\font plus
\BIBentryALTinterwordstretchfactor\fontdimen3\font minus
  \fontdimen4\font\relax}
\providecommand{\BIBforeignlanguage}[2]{{%
\expandafter\ifx\csname l@#1\endcsname\relax
\typeout{** WARNING: IEEEtran.bst: No hyphenation pattern has been}%
\typeout{** loaded for the language `#1'. Using the pattern for}%
\typeout{** the default language instead.}%
\else
\language=\csname l@#1\endcsname
\fi
#2}}
\providecommand{\BIBdecl}{\relax}
\BIBdecl

\bibitem{zhang2025vehicle}
X.~Zhang \emph{et~al.}, ``{Vehicle-to-Everything Communication in Intelligent
  Connected Vehicles: A Survey and Taxonomy},'' \emph{Automotive Innovation},
  pp. 1--33, 2025.

\bibitem{nguyen2025survey}
H.~Nguyen \emph{et~al.}, ``{Survey of Next-generation Optical Wireless
  Communication Technologies for 6G and Beyond 6G},'' \emph{ICT Express}, 2025.

\bibitem{decarli2023v2x}
N.~Decarli \emph{et~al.}, ``{V2X sidelink localization of connected automated
  vehicles},'' \emph{IEEE Journal on Selected Areas in Communications},
  vol.~42, no.~1, pp. 120--133, 2023.

\bibitem{cai2023consensus}
K.~Cai \emph{et~al.}, ``Consensus-based distributed cooperative perception for
  connected and automated vehicles,'' \emph{IEEE Transactions on Intelligent
  Transportation Systems}, vol.~24, no.~8, pp. 8188--8208, 2023.

\bibitem{barbieri2024lidar}
L.~Barbieri \emph{et~al.}, ``Deep learning-based cooperative lidar sensing for
  improved vehicle positioning,'' \emph{IEEE Transactions on Signal
  Processing}, vol.~72, pp. 1666--1682, 2024.

\bibitem{Memedi2021VVLC}
A.~Memedi and F.~Dressler, ``Vehicular visible light communications: A
  survey,'' \emph{IEEE Communications Surveys \& Tutorials}, vol.~23, no.~1,
  pp. 239--260, 2021.

\bibitem{Liu2020ISAC}
F.~Liu \emph{et~al.}, ``Toward dual-functional radar-communication systems:
  Optimal waveform design,'' \emph{IEEE Transactions on Signal Processing},
  vol.~66, no.~16, pp. 4264--4279, 2018.

\bibitem{Cheng2022ISACVehicular}
X.~Cheng \emph{et~al.}, ``{Integrated sensing and communications (ISAC) for
  vehicular communication networks (VCN)},'' \emph{IEEE Internet of Things
  Journal}, vol.~9, no.~23, pp. 23\,441--23\,451, 2022.

\bibitem{zhang2024lirf}
R.~Zhang \emph{et~al.}, ``{LiRF: Light-based Wireless Communications Supporting
  Ubiquitous Radio Frequency Signals},'' \emph{IEEE Photonics Journal}, 2024.

\bibitem{Dabiri2022MRRFSO}
M.~T. Dabiri \emph{et~al.}, ``{Modulating retroreflector based free space
  optical link for UAV-to-ground communications},'' \emph{IEEE Transactions on
  Wireless Communications}, vol.~21, no.~10, pp. 8631--8645, 2022.

\bibitem{10897286}
M.~T. Dabiri and M.~Hasna, ``{A Novel MRR-UAV-Based Relay With Optical Network
  Coding: A Comparative Study With Optical IRS and Conventional UAV
  Relaying},'' \emph{IEEE Journal on Selected Areas in Communications},
  vol.~43, no.~5, pp. 1607--1620, 2025.

\bibitem{chen2025line}
J.~Chen \emph{et~al.}, ``Line-structured light-based three-dimensional
  reconstruction measurement system with an improved scanning-direction
  calibration method,'' \emph{Remote Sensing}, vol.~17, no.~13, p. 2236, 2025.

\bibitem{ergin2024robotic}
E.~Ergin \emph{et~al.}, ``{Robotic IoT-enabled 1d line scanner integration for
  3D point cloud data processing},'' in \emph{Proceedings of the 41st
  International Symposium on Automation and Robotics in Construction}, 2024,
  pp. 1168--1175.

\bibitem{cai2008paraxial}
Y.~Cai \emph{et~al.}, ``{Paraxial propagation of a partially coherent flattened
  Gaussian beam through apertured ABCD optical systems},'' \emph{Optics
  communications}, vol. 281, no.~12, pp. 3221--3229, 2008.

\end{thebibliography}

\end{document}